\documentclass[aps,prl,twocolumn,amsmath,amssymb,superscriptaddress]{revtex4-1}

\usepackage{graphics}
\usepackage{dcolumn}
\usepackage{bm}
\usepackage[caption=false]{subfig} 
\usepackage{color}
\usepackage{tikz}
\usepackage{subfig}
\usepackage[titles,subfigure]{tocloft}
\usepackage{soul}

\usepackage[titletoc,toc,title]{appendix}

\begin{document}

\title{Floquet Topological Semimetal with Nodal Helix}

\author{Kun Woo Kim}
\affiliation{School of Physics, Korea Institute for Advanced Study, Seoul 02455, Korea}
\author{Hyun Woong Kwon}
\affiliation{Quantum Universe Center, Korea Institute for Advanced Study, Seoul 02455, Korea}
\author{Kwon Park}
\email[Electronic address:$~~$]{kpark@kias.re.kr}
\affiliation{School of Physics, Korea Institute for Advanced Study, Seoul 02455, Korea}
\affiliation{Quantum Universe Center, Korea Institute for Advanced Study, Seoul 02455, Korea}

\date{\today}

\begin{abstract}
Topological semimetals with nodal line are a novel class of topological matter extending the concept of topological matter beyond topological insulators and Weyl/Dirac semimetals.  
Here, we show that a Floquet topological semimetal with nodal helix can be generated by irradiating graphene or the surface of a topological insulator with circularly polarized light.
Nodal helix is a form of nodal line running across the Brillouin zone with helical winding.
Specifically, it is shown that the dynamics of irradiated graphene is described by the time Stark Hamiltonian, which can host a Floquet topological insulator and a weakly driven Floquet topological semimetal with nodal helix in the high and low frequency limits, respectively. 
It is predicted that, at low frequency, the $\pi$ shift of the Zak phase generates a topological discontinuity along the projected nodal helix in the momentum spectrum of the Floquet states.
At intermediate frequency, this topological discontinuity can create an interesting change of patterns in the quasienergy dispersion of the Floquet states.  
\end{abstract}

\maketitle

Topological matter can be classified in various ways.  
An intuitive way is to note how Dirac monopoles are located in the Hamiltonian parameter space. 
In the case of topological insulators~\cite{Haldane1988, Kane2005, Konig2007, Hasan2010, Qi2011}, Dirac monopoles exist in an appropriate, but invisible parameter space, while avoided in the momentum space.
In the case of Weyl~\cite{Murakami2007, Wan2011} and Dirac~\cite{Wang2012} semimetals, they are directly located in the momentum space as isolated points.
Topological semimetals with nodal line~\cite{Burkov2011} can be regarded as a novel class of topological matter, where Dirac monopoles form a closed loop in the momentum space. 
Despite considerable attention, however, conclusive experimental evidence for their existence has been elusive so far partly since they require a rather delicate symmetry protection in real materials~\cite{Xu2011, Carter2012, Weng2015, Fang2015, Kim2015, Yu2015, Xie2015, Chen2015, Liang2016, Huang2016, Chan2016, Yamakage2016}.

Meanwhile, there has been a rapidly growing interest in the artificial generation of topological matter dubbed as Floquet engineering~\cite{Oka2018}.
One of the most notable examples is the theoretical proposal for the generation of a Floquet topological insulator by irradiating graphene~\cite{Oka2009, Kitagawa2011, Kundu2014, Dehghani2015, Sentef2015, Mikami2016} or semiconductor quantum wells~\cite{Lindner2011}.
The Floquet topological insulator generated in the high frequency limit of irradiated graphene is particularly interesting since it can provide an exact realization of the Haldane model~\cite{Haldane1988} or the Kane-Mele model~\cite{Kane2005} for a single spin species with the possibility of manipulating the Chern number via tuning the radiation frequency and electric field strength.

Here, we show that, in addition to the Floquet topological insulator in the high frequency limit, irradiated graphene can host a weakly driven Floquet topological semimetal with nodal helix at low frequency.  
With the nodal helix being a form of nodal line running across the time Brillouin zone~\cite{Chen2015}, this provides a novel platform for the artificial generation of topological semimetals with nodal line.  
A salient feature of the so-obtained Floquet topological semimetal is the $\pi$ shift of the Zak phase inside the projected nodal helix, giving rise to the topological discontinuity along the projected nodal helix in the momentum spectrum of the Floquet states. 
At intermediate frequency, this topological discontinuity can create an interesting change of patterns in the quasienergy dispersion of the Floquet states, which cannot be understood in terms of the simple overlapping Floquet copies of the Dirac dispersion.
Having the same dispersion as graphene in the continuum limit, the surface of a topological insulator can also serve as another promising platform for the Floquet topological semimetal with nodal helix.
We discuss the feasibility of its experimental observation via time- and angle-resolved photoemission spectroscopy (Tr-ARPES)~\cite{Wang2013, Mahmood2016}.

\begin{figure*}[t]
\begin{center}
\includegraphics[width=1.5\columnwidth]{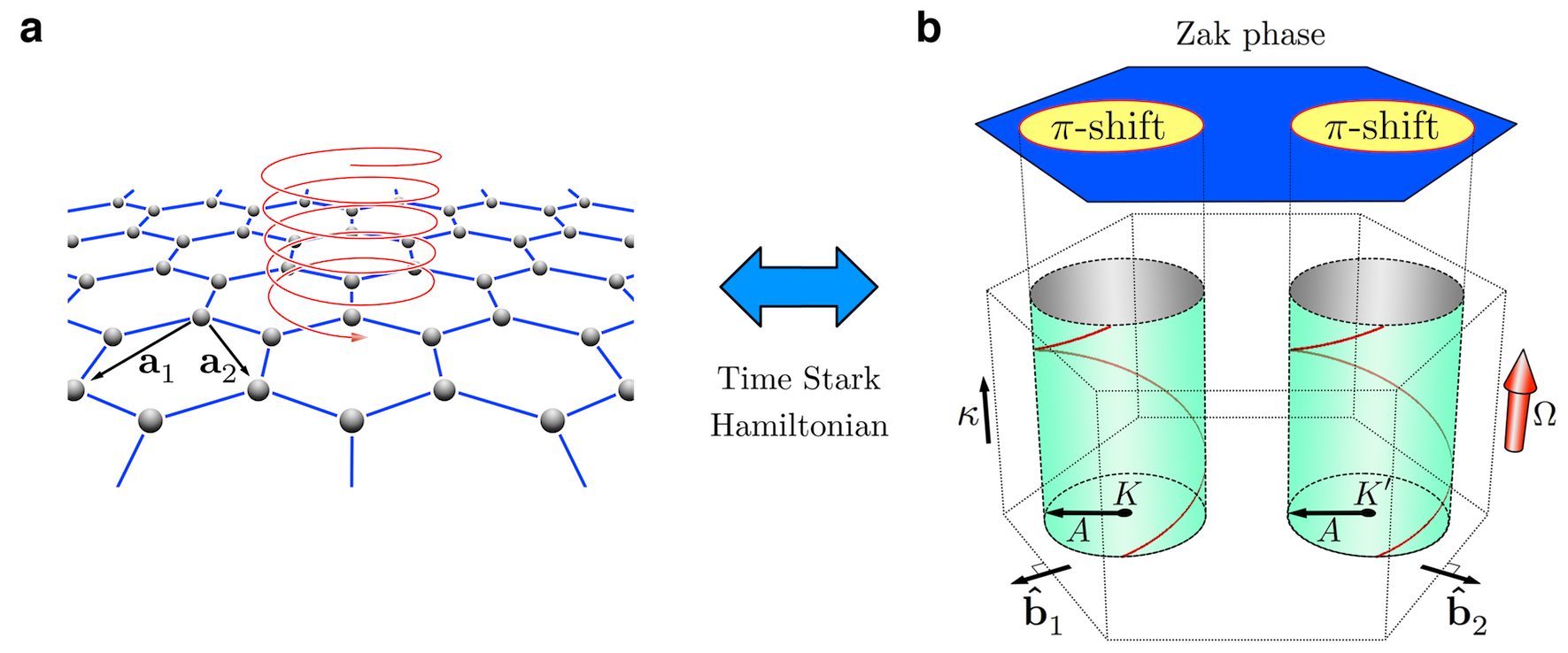}
\caption{
{\bf Schematic diagram showing the mapping of irradiated graphene from the real (a) to the extended momentum (b) space.} 
Here, ${\bf a}_1$ and ${\bf a}_2$ denote the primitive lattice vectors of graphene in the real space, while ${\bf \hat{b}}_1$ and ${\bf \hat{b}}_2$ denote the directions of their respective reciprocal lattice vectors in the extended momentum space.
The dynamics of irradiated graphene is described by the time Stark Hamiltonian in the extended momentum space, where an effective DC electric field with the strength equal to the radiation frequency $\Omega$ is applied along the axis of the nodal helix aligned with the time, i.e., $\kappa=\Omega t$ direction.
Note that the nodal helix is the trajectory of each Dirac node (centered at either $K$ or $K^\prime$) as a function of time.  
The radius of the nodal helix is given by $A=eE_0/\Omega$ with $E_0$ being the radiation electric field strength.
The Zak phase accumulated along the $\kappa$ direction acquires a relative $\pi$ shift inside the projected nodal helix due to the same reason why the Su-Schrieffer-Heeger model becomes topological.
At low frequency, the $\pi$ shift of the Zak phase gives rise to the topological discontinuity along the projected nodal helix in the momentum spectrum of the Floquet states.
\label{fig:Schematic}
}
\end{center}
\end{figure*}

{\bf Results}

{\bf Time Stark Hamiltonian.}
When the Hamiltonian $\hat{H}(t)$ is periodic in time with period $T=2\pi/\Omega$, the Floquet theorem dictates that the solution of the Schr\"{o}dinger equation can be generally written as $|\Psi_\alpha(t)\rangle = e^{-i\epsilon_\alpha t/\hbar} |\psi_\alpha(t)\rangle$ with $|\psi_\alpha(t)\rangle$ being periodic in time with the same period,
i.e., $|\psi_\alpha(t)\rangle=\sum_n  e^{-in\Omega t} |\tilde{\psi}_\alpha^n\rangle$, where $n$ is the Fourier index conjugate to time.
The quasienergy eigenvalue $\epsilon_\alpha$ and the time-conjugate Fourier components $\{|\tilde{\psi}_\alpha^n\rangle\}$ of the $\alpha$-th Floquet state can be determined by diagonalizing the Floquet Hamiltonian:
\begin{align}
[\hat{H}_{\rm F}]_{nm} = \hat{H}_{nm} +n\Omega \delta_{nm} ,
\label{eq:H_F}
\end{align}
where $\hat{H}_{nm}=\frac{1}{T} \int^T_0 \hat{H}(t) e^{i(n-m)\Omega t} dt$.
We set $\hbar=1$ from this foward.

The Floquet Hamiltonian can be interpreted as the Stark Hamiltonian in an effective parameter space composed of the momentum and time-conjugate Fourier indices, where an effective electric field with strength $\Omega$ is applied along the time-conjugate Fourier index, or simply the time direction~\cite{Gomez-Leon2013}.
In this work, we show that, to achieve a unified understanding of the two opposite limits of high and low frequencies, it is beneficial to work with the Stark Hamiltonian directly expressed in terms of time, which we call the time Stark Hamiltonian (TSH): 
\begin{align}
\hat{H}_{\rm TS}({\bf k})= \hat{\cal E}({\bf k}) +\Omega \left[ i\frac{\partial}{\partial \kappa} +\hat{\cal A}_\kappa({\bf k}) \right] ,
\label{eq:H_TS}
\end{align}
where ${\bf k}=({\bf k}_\perp, \kappa)$ is the extended momentum with two real, ${\bf k}_\perp=(k_x,k_y)$, and one effective, $\kappa=\Omega t$, components for the time-periodic 2D system, which is of our interest in this work.
$\hat{\cal E}({\bf k})$ is the {\it instantaneous} energy eigenvalue matrix obtained by diagonalizing the nominal Hamiltonian, $\hat{H}({\bf k})$, which is simply the original Hamiltonian with $\Omega t$ just replaced by $\kappa$. 
Mathematically, $[\hat{\cal E}({\bf k})]_{ab}=\epsilon_a({\bf k}) \delta_{ab}$ with $\epsilon_a({\bf k})$ being the instantaneous eigenvalue of $\hat{H}({\bf k})$ for the $a$-th band.  
$\hat{\cal A}_\kappa({\bf k})$ is the non-Abelian Berry connection projected along the $\kappa$ direction; $[\hat{\cal A}_\kappa({\bf k})]_{ab} \equiv {\cal A}_{\kappa}^{ab}({\bf k})=\langle u_a({\bf k})| i \frac{\partial}{\partial \kappa} | u_b({\bf k})\rangle$ with $|u_a({\bf k})\rangle$ and $|u_b({\bf k})\rangle$  being the instantaneous eigenstates of $\hat{H}({\bf k})$ for the $a$-th and $b$-th bands, respectively.
Note that essentially the same Hamiltonian has been studied in the context of topological insulators~\cite{Lee2015} and Weyl semimetals~\cite{Kim2016} under a real DC electric field.

To be specific, the graphene Hamiltonian without radiation can be written as follows:
\begin{align}
\hat{H}_0({\bf k}_\perp)=
\left(
\begin{array}{cc}
0 & g_{{\bf k}_\perp} \\
g^*_{{\bf k}_\perp} & 0
\end{array}
\right) ,
\label{eq:H_0}
\end{align} 
where $g_{{\bf k}_\perp}=-\tau(e^{i{\bf k}_\perp \cdot {\bf c}_1}+e^{i{\bf k}_\perp \cdot {\bf c}_2}+e^{i{\bf k}_\perp \cdot {\bf c}_3})$ with ${\bf c}_1=(-\sqrt{3}/2, -1/2)$, ${\bf c}_2=(\sqrt{3}/2, -1/2)$, and ${\bf c}_3=(0,1)$ in units of lattice spacing, which is set to be unity throughout this paper.
$\tau$ is the hopping parameter between nearest neighboring sites.
The Hamiltonian in the presence of circularly polarized radiation can be obtained via Peierls substitution, ${\bf k}_\perp \rightarrow {\bf k}_\perp^{\rm Pei\mbox{-}sub}={\bf k}_\perp - \frac{e}{c}{\bf A}(t)$, where the vector potential is ${\bf A}(t)=\frac{c E_0}{\Omega}(\cos{\Omega t},\sin{\Omega t})$ with $E_0$ being the radiation electric field strength. 
In the extended momentum notation, ${\bf k}_\perp^{\rm Pei\mbox{-}sub}=(k_x-A\cos{\kappa},k_y-A\sin{\kappa})$ with $A=e E_0/\Omega$. 
Consequently, the graphene Hamiltonian in the presence of circularly polarized radiation is written in the extended momentum notation as follows:
\begin{align}
\hat{H}({\bf k})=
\left(
\begin{array}{cc}
0 & g_{\bf k} \\
g^*_{\bf k} & 0
\end{array}
\right) ,
\label{eq:H_k}
\end{align} 
where $g_{\bf k}=g_{{\bf k}_\perp^{\rm Pei\mbox{-}sub}}$.

The instantaneous eigenstates of $\hat{H}({\bf k})$ in Eq.~\eqref{eq:H_k} are given by
\begin{align}
|u_{\pm}({\bf k})\rangle = \frac{1}{\sqrt{2}}
\left(
\begin{array}{c}
e^{-i\phi({\bf k})} \\
\pm 1
\end{array}
\right) ,
\label{eq:insta_eigenstates}
\end{align}
where $e^{-i\phi({\bf k})}=g_{\bf k}/|g_{\bf k}|$.
The corresponding instantaneous eigenvalues are $\epsilon_{\pm}({\bf k})=\pm|g_{\bf k}|$, where $\pm$ indicates the conduction and valence bands, respectively.
That is,
\begin{align}
\hat{\cal E} ({\bf k}) = 
\left(
\begin{array}{cc}
|g_{\bf k}| & 0\\
0 & -|g_{\bf k}|
\end{array}
\right) ,
\label{eq:E_hat}
\end{align}
meaning that the energy gap closes when $|g_{\bf k}|=0$. 
Visually, each Dirac node spirals around its original position, forming a nodal helix with radius $A$.
Meanwhile, the instantaneous eigenstates give rise to the following Berry connection: 
\begin{align}
\hat{\cal A}_\kappa({\bf k}) = 
\frac{1}{2} \frac{\partial \phi({\bf k})}{\partial \kappa}
\left(
\begin{array}{cc}
1 & 1\\
1 & 1
\end{array}
\right) .
\label{eq:Berry_connection}
\end{align}
In summary, the dynamics of irradiated graphene is described by the TSH, where an effective DC electric field is applied along the axis of the nodal helix.
See Fig.~\ref{fig:Schematic} for illustration.

The TSH needs to be solved numerically at general frequency.
Fortunately, however, one can obtain quite useful analytical expressions for the two topologically interesting solutions emerging in the respective limits of high and low frequencies.  
The main goal of this work is to investigate what happens in the low frequency limit.
Before elaborating on this, it is instructive to show how the TSH can capture the emergence of a Floquet topological insulator in the high frequency limit.

{\bf Floquet topological insulator at high frequency.}
It is well known that a Floquet topological insulator can emerge in the high frequency limit of irradiated graphene with circularly polarized light~\cite{Oka2009, Kitagawa2011, Kundu2014, Dehghani2015, Sentef2015, Mikami2016}.
A question is how exactly the TSH in Eq.~\eqref{eq:H_TS} can capture this.

In the high frequency limit, one can proceed by first solving the second term in Eq.\eqref{eq:H_TS}, which has $\Omega$ as a prefactor, 
\begin{align}
\frac{H_\Omega}{\Omega} = i\frac{\partial}{\partial \kappa} 
+\frac{1}{2} \frac{\partial \phi({\bf k})}{\partial \kappa}
\left(
\begin{array}{cc}
1 & 1\\
1 & 1
\end{array}
\right) ,
\label{eq:H_Omega}
\end{align}
and then taking into account the first term, $\hat{\cal E}$.
 $H_\Omega$ can be diagonalized analytically by the following eigenstates:
\begin{align}
|\varphi_{\pm,\nu}({\bf k})\rangle =\frac{1}{\sqrt{2}}
\left(
\begin{array}{c}
1 \\
\pm 1
\end{array}
\right)
f_{\pm,\nu} ({\bf k}) ,
\label{eq:high_freq_eigenstates}
\end{align}
where $f_{+,\nu}({\bf k})=e^{-i [\nu \kappa-\Delta\phi({\bf k})]}$ and $f_{-,\nu}({\bf k})=e^{-i \nu \kappa}$ with $\Delta\phi({\bf k})=\phi({\bf k}_\perp,\kappa)-\phi({\bf k}_\perp,0)$.
Due to the periodic boundary condition, the eigenvalue $\nu$ is an integer.

By using $\{|\varphi_{\pm,\nu}({\bf k})\rangle\}$ as a new set of basis, one can express $\hat{\cal E}$ in the following matrix form:
\begin{align}
{\cal M}_{\hat{\cal E}} =
\left(
\begin{array}{ccccc}
\ddots & \ddots & \ddots & &  \\
\ddots & \hat{\Gamma}_{0} & \hat{\Gamma}_1^\dagger  & \hat{\Gamma}_2^\dagger & \\
\ddots & \hat{\Gamma}_1 & \hat{\Gamma}_0      & \hat{\Gamma}_1^\dagger & \ddots \\
 & \hat{\Gamma}_2 & \hat{\Gamma}_1  & \hat{\Gamma}_0 & \ddots \\
 & & \ddots & \ddots & \ddots \\
\end{array}
\right) ,
\label{eq:M_E}
\end{align}
where 
\begin{align}
\hat{\Gamma}_n =
\left(
\begin{array}{cc}
0 & \Gamma_n({\bf k}_\perp) \\
\Gamma^*_{-n}({\bf k}_\perp) & 0
\end{array}
\right) ,
\label{eq:Gamma_matrix}
\end{align}
and $\Gamma_n({\bf k}_\perp)=e^{i\phi_0({\bf k}_\perp)} \int_0^{2\pi} \frac{d\kappa}{2\pi} g_{\bf k} e^{i n\kappa}$ with $\phi_0({\bf k}_\perp)=\phi({\bf k}_\perp,\kappa=0)$.
The entire TSH including both terms in Eq~\eqref{eq:H_TS} can be written as ${\cal H}_{\rm TS} = {\cal M}_{\hat{\cal E}} +{\cal M}_\Omega$, where ${\cal M}_\Omega$ is the diagonal matrix with $[{\cal M}_\Omega]_{\nu \nu^\prime}=\nu \Omega \hat{I} \delta_{\nu \nu^\prime}$ for the $(\nu,\nu^\prime)$-th $2 \times 2$ block.

Now, ${\cal H}_{\rm TS}$ can be systematically expanded as a power series of $1/\Omega$. 
Specifically, valid up to the order of $1/\Omega$, the following effective $2 \times 2$ Hamiltonian can be obtained for the $\nu$-th pair of the quasienergy bands:
\begin{align}
{\cal H}_{{\rm eff},\nu} =
\left(
\begin{array}{cc}
\nu\Omega & \Gamma_0 \\
\Gamma_0^* & \nu\Omega
\end{array}
\right)
+\sum_{n=1}^\infty \frac{|\Gamma_n|^2-|\Gamma_{-n}|^2}{n\Omega}
\left(
\begin{array}{cc}
1 & 0 \\
0 & -1
\end{array}
\right) .
\label{eq:H_eff}
\end{align}
Note that the above effective Hamiltonian can be derived by using the combination of the usual degenerate perturbation theory treating the effects of $\Gamma_0$ and the second-order virtual process treating those of $\Gamma_{n \neq 0}$.

After performing a series of integrations and algebras, one can show that, in the high frequency limit, ${\cal H}_{{\rm eff},\nu}$ can be rewritten as follows:
\begin{align}
{\cal H}_{{\rm eff},\nu}= \nu\Omega \hat{I} +{\bf d}_{{\bf k}_\perp} \cdot \boldsymbol{\sigma} ,
\label{eq:H_eff_main}
\end{align}
where 
$d^*_{{\bf k}_\perp,+}=d_{{\bf k}_\perp,-} = - \tau_{\rm eff}(A) e^{i\phi_0({\bf k}_\perp)} \sum_{j=1}^3 e^{i{\bf k}_\perp \cdot {\bf c}_j}$
with $d_{{\bf k}_\perp,\pm}=d_{{\bf k}_\perp,x} \pm i d_{{\bf k}_\perp,y}$, and
$d_{{\bf k}_\perp,z} = i \lambda_{\rm eff}(A) \sum_{j=1}^6 (-1)^j e^{i {\bf k}_\perp \cdot \boldsymbol{\eta}_j}$ 
with $\boldsymbol{\eta}_j$ being the displacement vectors connecting between next-nearest-neighboring sites: $\boldsymbol{\eta}_1 = (-\sqrt{3}/2, -3/2)$, $\boldsymbol{\eta}_2 = (\sqrt{3}/2, -3/2)$, $\boldsymbol{\eta}_3 = (\sqrt{3},0)$,  $\boldsymbol{\eta}_4 = (\sqrt{3}/2, 3/2)$, $\boldsymbol{\eta}_5 = (-\sqrt{3}/2, 3/2)$, and $\boldsymbol{\eta}_6 = (-\sqrt{3},0)$.
The effective hopping and spin-orbit coupling parameters are given as $\tau_{\rm eff}(A)=\tau J_0(A)$ and $\lambda_{\rm eff}(A) = \frac{2 \tau^2}{\Omega} \sum_{n=1}^\infty \frac{J_n^2(A)}{n} \sin{\left( \frac{2n\pi}{3} \right)}$, respectively.
It is important to note that these formulas are exactly the same as those previously obtained by using the Brillouin-Wigner theory~\cite{Mikami2016}.
See {\bf Methods} for the details of the derivation.

In summary, up to the order of $1/\Omega$ in the high frequency limit, the TSH can be decomposed into an infinite ladder of the effective $2 \times 2$ Hamiltonians, ${\cal H}_{{\rm eff},\nu}$, each of which is exactly identical to the Kane-Mele model Hamiltonian~\cite{Kane2005} of a single spin species with the effective hopping parameter $\tau_{\rm eff}(A)$ and the effective spin-orbit coupling parameter $\lambda_{\rm eff}(A)$. 
Since the Kane-Mele model is nothing but two copies of the Haldane model with opposite flux configurations for different spin species, this means that ${\cal H}_{{\rm eff},\nu}$ is also exactly identical to the Haldane model Hamiltonian~\cite{Haldane1988} describing a Chern insulator.

\begin{figure*}[t]
\begin{center}
\includegraphics[width=1.1\columnwidth]{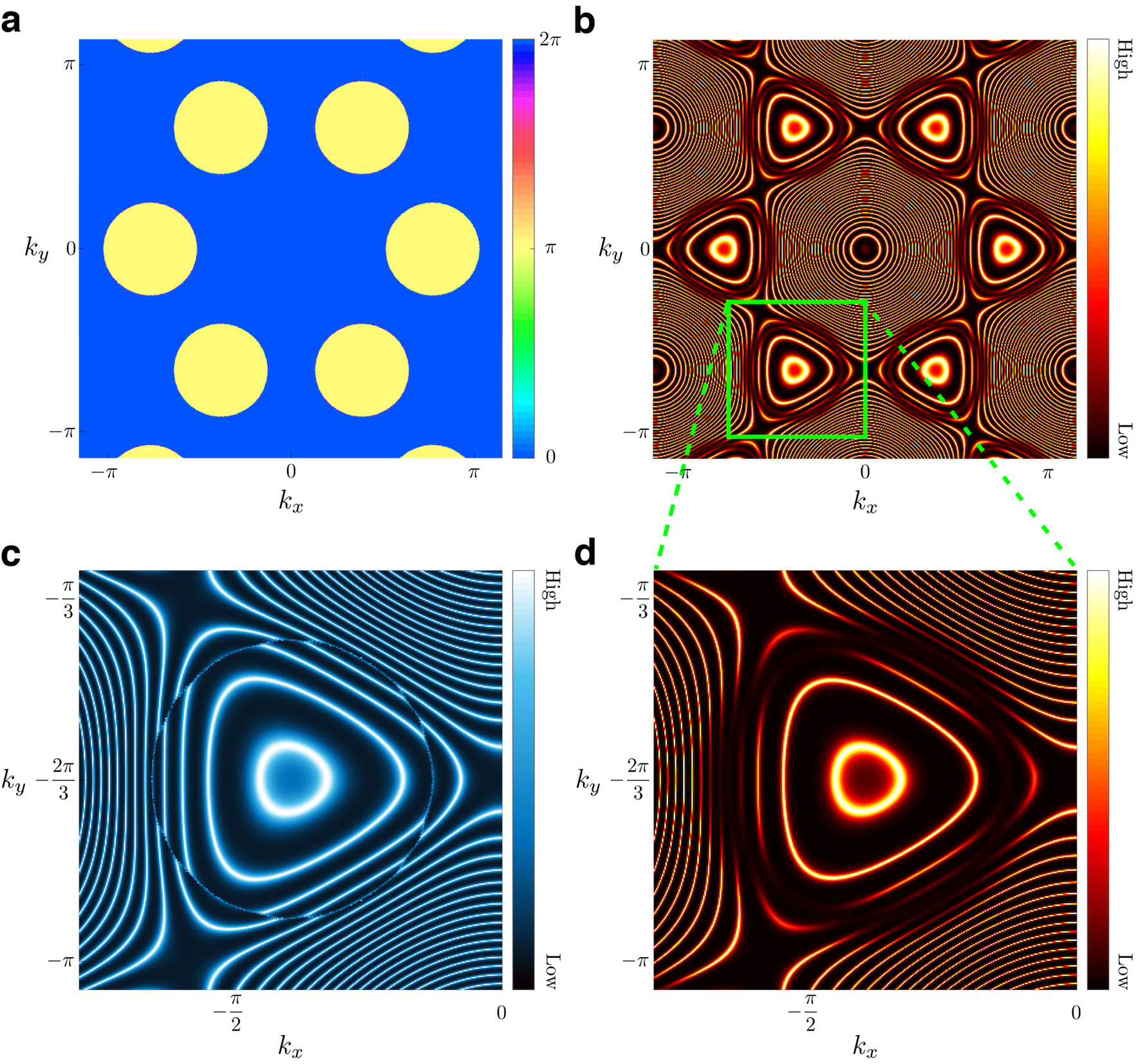}
\caption{
{\bf Zero-energy momentum spectrum of the Floquet states in irradiated graphene at low frequency.}
Here, the radiation frequency is set to be $\Omega/\tau=0.05$ with $\tau$ being the hopping parameter, and the radius of the nodal helix is set to be $A=0.8$. 
({\bf a}) Zak phase accumulated along the time direction, showing the $\pi$ shift inside the projected nodal helix.
({\bf b}) Zero-energy momentum spectrum of the full Floquet states obtained by exactly solving either the Floquet or the time Stark Hamiltonian.
({\bf c}) Closeup of the zero-energy momentum spectrum of the time Wannier-Stark ladder eigenstates obtained by solving the Abelian approximation of the time Stark Hamiltonian. 
({\bf d}) Closeup of the zero-energy momentum spectrum of the full Floquet states in the same region as ({\bf c}), denoted as the green box in ({\bf b}).
\label{fig:Topological_discontinuity}
}
\end{center}
\end{figure*}

{\bf Weakly driven Floquet topological semimetal with nodal helix at low frequency.}
The low frequency limit of the TSH is particularly interesting in the context of topological semimetals with nodal line.  
In the low frequency limit, the band-mixing, i.e., off-diagonal elements of the non-Abelian Berry connection can be ignored.
That is, the Berry connection can be approximated to be Abelian.
The eigenstates of the Abelian Stark Hamiltonian under a DC electric field are known as the Wannier-Stark ladder (WSL) eigenstates. 
By analogy, let us call the eigenstates of the Abelian TSH the time WSL eigenstates.

The Abelian TSH for the $a$-th band can be written as 
\begin{align}
\hat{H}_{{\rm ATS},a} = \epsilon_a({\bf k}) +\Omega \left( i\frac{\partial}{\partial \kappa}  +{\cal A}^{aa}_\kappa({\bf k}) \right) ,
\end{align}
which can be in turn diagonalized by the following time WSL eigenstates:
\begin{align}
\psi^{\rm WSL}_{a,n} ({\bf k}) = e^{ 
-\frac{i}{\Omega} \int^\kappa_0 d\kappa^\prime 
\left[ {\cal E}^{\rm WSL}_{a,n} ({\bf k}_\perp) -\epsilon_a({\bf k}^\prime) -\Omega {\cal A}^{aa}_\kappa({\bf k}^\prime) \right]  
} ,
\label{eq:psi_WSL}
\end{align}
where ${\bf k}^\prime=({\bf k}_\perp,\kappa^\prime)$.
The quasienergy of the time WSL eigenstates is determined by the periodic boundary condition;
\begin{align}
{\cal E}^{\rm WSL}_{a,n}({\bf k}_\perp) = \bar{\epsilon}_a({\bf k}_\perp) +\Omega \left(n+\frac{\gamma^{\rm Zak}_a({\bf k}_\perp)}{2\pi}\right) ,
\label{eq:E_WSL}
\end{align}
where the time WSL index $n$ is an integer, $\bar{\epsilon}_a({\bf k}_\perp)=\frac{1}{2\pi}\oint d\kappa \epsilon_a({\bf k})$ is the time-averaged instantaneous band energy, and 
$\gamma^{\rm Zak}_a({\bf k}_\perp)=\oint d\kappa {\cal A}^{aa}_\kappa({\bf k})$ is the Zak phase accumulated along the $\kappa$ direction.

\begin{figure*}[t]
\begin{center}
\includegraphics[width=1.3\columnwidth]{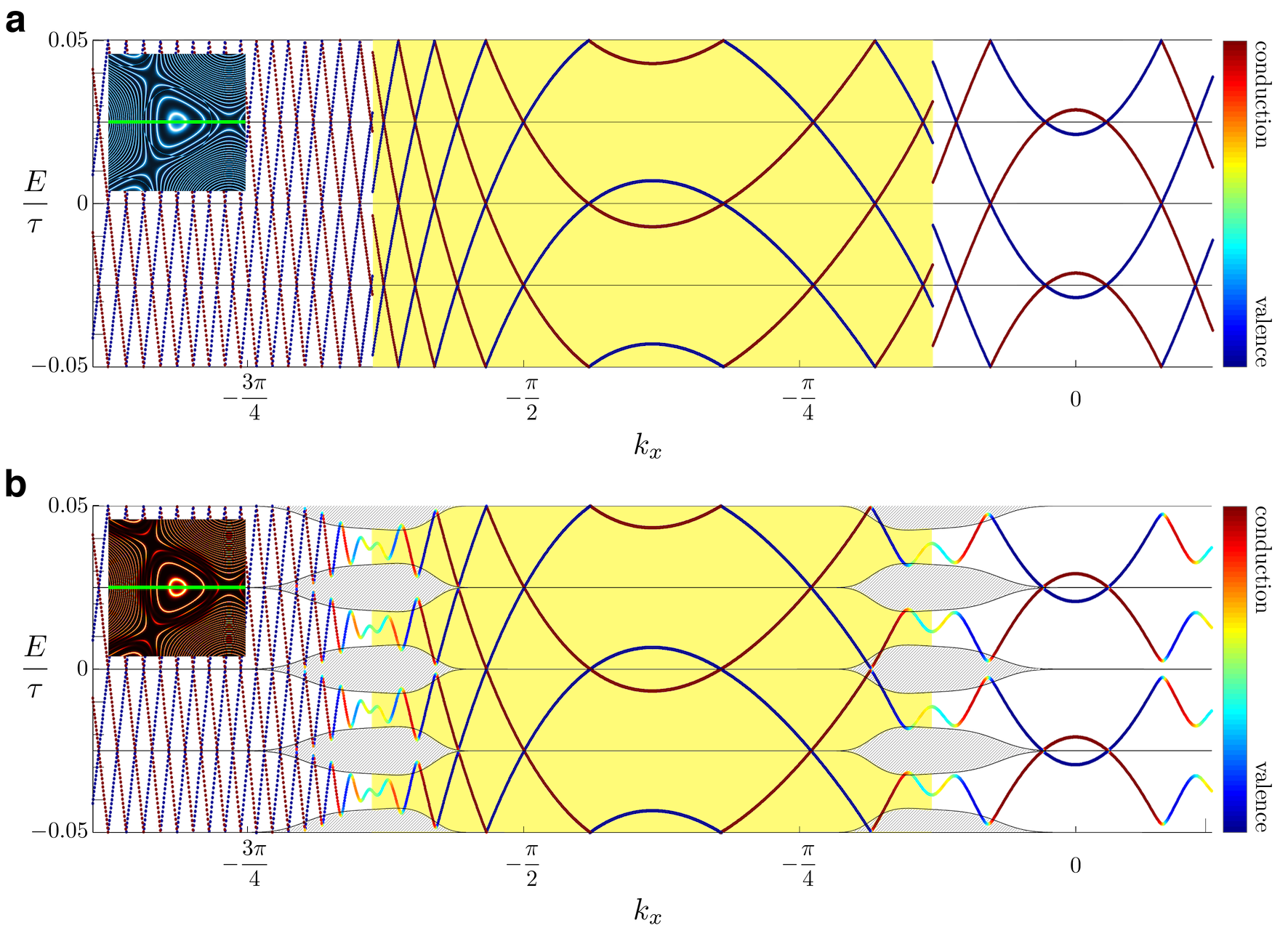}
\caption{
{\bf Quasienergy dispersion of the Floquet states in irradiated graphene.}
Here, the radiation frequency and the radius of the nodal helix are set to be the same corresponding values as Fig.~\ref{fig:Topological_discontinuity}.
({\bf a}) Quasienergy dispersion of the time Wannier-Stark ladder eigenstates showing the mismatch of the quasienergy eigenvalues and hence the topological discontinuity at the projected nodal helix.
Note that the inside of the projected nodal helix is denoted as a yellow region.
({\bf b}) Quasienergy dispersion of the full Floquet states showing the opening of the quasienergy gap near the projected nodal helix.
The quasienergy gap is very well captured by the analytical formula ${\cal E}_{\rm gap}$ in Eq.~\eqref{eq:E_gap_main}, the result of which is plotted as the shaded (grey) regions bounded by the thin black lines in the figure.
Note that the thin black lines denote the plus and minus halves of ${\cal E}_{\rm gap}$. 
Colour indicates the Abelian weight measuring how much the Floquet states belongs to the conduction (red) or valence (blue) band. 
By construction, the time Wannier-Stark ladder eigenstates belong to either one of the two bands exclusively. 
The green lines in the insets show the scanned momentum path, along which the quasienergy dispersion is computed. 
\label{fig:Quasienergy_gap}
}
\end{center}
\end{figure*}

The time WSL eigenstates of irradiated graphene have an intriguing topological structure induced by the $\pi$ shift of the Zak phase inside the projected nodal helix. 
Mathematically, $\gamma^{\rm Zak}_\pm({\bf k}_\perp)=\frac{1}{2} \oint d \kappa \frac{\partial}{\partial \kappa} \phi({\bf k})=\pi$ and 0 for ${\bf k}_\perp$ located inside and outside the projected nodal helix, respectively.
The reason is actually exactly the same as why the Su-Schrieffer-Heeger model becomes topological~\cite{Su1980, Zak1989, King-Smith1993, Delplace2011}. 
Consequently, the quasienergy of the time WSL eigenstates acquires the relative $\Omega/2$ shift inside the projected nodal helix:
\begin{align}
{\cal E}^{\rm WSL}_{\pm,n}({\bf k}_\perp) = 
\Big\{
\begin{array}{ll}
\bar{\epsilon}_\pm({\bf k}_\perp) +\Omega (n+1/2) &  {\rm (inside)} \\
\bar{\epsilon}_\pm({\bf k}_\perp) +\Omega n          &  {\rm (outside)}
\end{array} ,
\label{eq:E_WSL2}
\end{align}
giving rise to the topological discontinuity along the projected nodal helix.

Figure~\ref{fig:Topological_discontinuity} shows the zero-energy momentum spectrum of the Floquet states in irradiated graphene at low frequency showing the topological discontinuity along the projected nodal helix.
The zero-energy momentum spectrum of the time WSL eigenstates can be obtained by tracking all the ${\bf k}_\perp$ curves satisfying ${\cal E}^{\rm WSL}_{\pm,n}({\bf k}_\perp)=0$ for different $n$.
Meanwhile, the zero-energy momentum spectrum  of the full Floquet states can be obtained by exactly solving either the Floquet Hamiltonian or the TSH.
Specifically, the quasienergy spectral function of the full Floquet states can be computed as
\begin{align}
\rho(\omega,{\bf k}_\perp) = -\frac{1}{\pi} {\rm Im} {\rm Tr} 
\left[
\frac{1}{\omega-\hat{H}_{\rm F}+i\eta} 
\right] ,
\label{eq:spectral_func}
\end{align}
where $\hat{H}_{\rm F}$ is the Floquet matrix in Eq.~\eqref{eq:H_F} and the trace Tr is taken over both Floquet and sublattice indices.
Then, the zero-energy momentum spectrum of the full Floquet states can be obtained by plotting $\rho(\omega=0,{\bf k}_\perp)$ as a function of ${\bf k}_\perp$. 
Alternatively, one can just directly diagonalize the TSH by using the method developed in a previous work by some of the current authors~\cite{Kim2016}. 
A main message of Fig.~\ref{fig:Topological_discontinuity} is that the zero-energy momentum spectrum of the full Floquet states is overall quite well captured by that of the time WSL eigenstates, providing a natural explanation for the existence of the topological discontinuity along the projected nodal helix in terms of the $\pi$ shift of the Zak phase.

Strictly speaking, however, the time WSL eigenstates should be valid only in the limit of low frequency.
Away from this limit, there is generally a mixing between different time WSL eigenstates from the conduction ($+$) and valence ($-$) bands, which opens up the quasienergy gap near the projected nodal helix, smearing out the topological discontinuity. 
Fortunately, it can be shown that the quasienergy gap is highly localized near the projected nodal helix. 
The quasienergy gap can be computed quite accurately near the projected nodal helix by using some form of the saddle point approximation.
In particular, a quite accurate analytical formula can be obtained in the continuum limit, where the graphene Hamiltonian is taken to be linear in the vicinity of the Dirac node.
Concretely, the quasienergy gap near the projected nodal helix is given by
\begin{align}
{\cal E}_{\rm gap} 
\simeq
\frac{\Omega A}{\sqrt{2\pi\rho(A+\rho)}}
e^{-\frac{2(A-\rho)^4}{\rho(A+\rho)(\Omega/\tau)^2}  } ,
\label{eq:E_gap_main}
\end{align}
where $\rho=|{\bf k}_\perp-{\bf K}_{\rm Dirac}|$ is the distance from the Dirac node.
See {\bf Methods} for the details of the derivation.
It is important to note that the quasienergy gap vanishes much faster than $\Omega$ as soon as $\rho$ becomes separated from $A$ by roughly more than $\sqrt{A\Omega/\tau}$.
This means that the topological discontinuity along the projected nodal helix becomes more and more pronounced in the low frequency limit.

\begin{figure*}[t]
\begin{center}
\includegraphics[width=1.7\columnwidth]{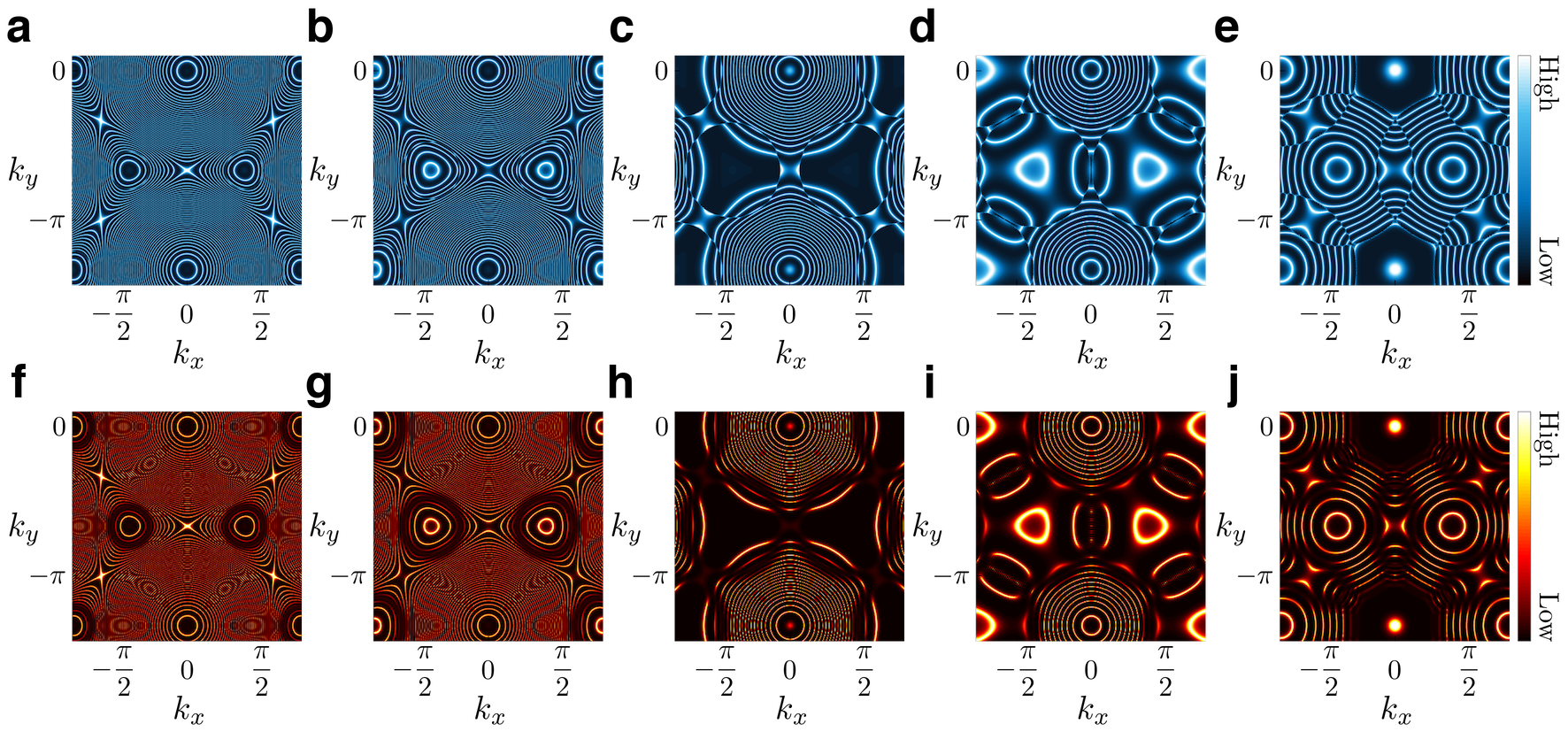}
\caption{
{\bf Evolution of the Floquet states in irradiated graphene as a function of the radius of the nodal helix.}
({\bf a}\mbox{-}{\bf e}) Zero-energy momentum spectrum of the time Wannier-Stark ladder eigenstates.
({\bf f}\mbox{-}{\bf j}) Zero-energy momentum spectrum of the full Floquet states.
Here, the radius of the nodal helix $A$ is changed from 0.4 ({\bf a}, {\bf f}) to 0.6 ({\bf b}, {\bf g}) to 1.0 ({\bf c}, {\bf h}) to 1.2 ({\bf d}, {\bf i}) to 1.6 ({\bf e}, {\bf j}), while the radiation frequency $\Omega/\tau$ is fixed as 0.05.
Note that the projected nodal helices generated from different Dirac nodes are just about to touch at ({\bf d}, {\bf i}) and pass through each other at ({\bf e}, {\bf j}).
\label{fig:Evolution}
}
\end{center}
\end{figure*}

Figure~\ref{fig:Quasienergy_gap} shows the comparison between the quasienergy dispersions of the time WSL eigenstates and the full Floquet states.
It is important to note that the quasienergy gap is very well described by the analytical formula ${\cal E}_{\rm gap}$ in Eq.~\eqref{eq:E_gap_main}.
This means that the full Floquet states are essentially given by the time WSL eigenstates except for the immediate vicinity of the projected nodal helix. 
This fact is reaffirmed by the Abelian weight measuring how much the Floquet states belongs to the conduction or valence band, or equivalently the overlap between the time WSL eigenstates and the full Floquet states.
See {\bf Methods} for the details on how to compute the Abelian weight.

Figure~\ref{fig:Evolution} shows the evolution of the Floquet states as a function of the radius of the nodal helix.
As one can see, the time Wannier-Stark ladder eigenstates can capture the evolution of the full Floquet states very well in a wide range of the radius of the nodal helix.
It is important to note that the momentum spectrum undergoes an interesting transition when the projected nodal helices generated from different Dirac nodes pass through each other.

{\bf Quasienergy dispersion at intermediate frequency.}
Generally, the quasienergy dispersion is quite complicated at intermediate frequency.  
Nevertheless, the time WSL eigenstates can provide a useful guide to the full quasienergy dispersion.

\begin{figure*}[t]
\begin{center}
\includegraphics[width=1.5\columnwidth]{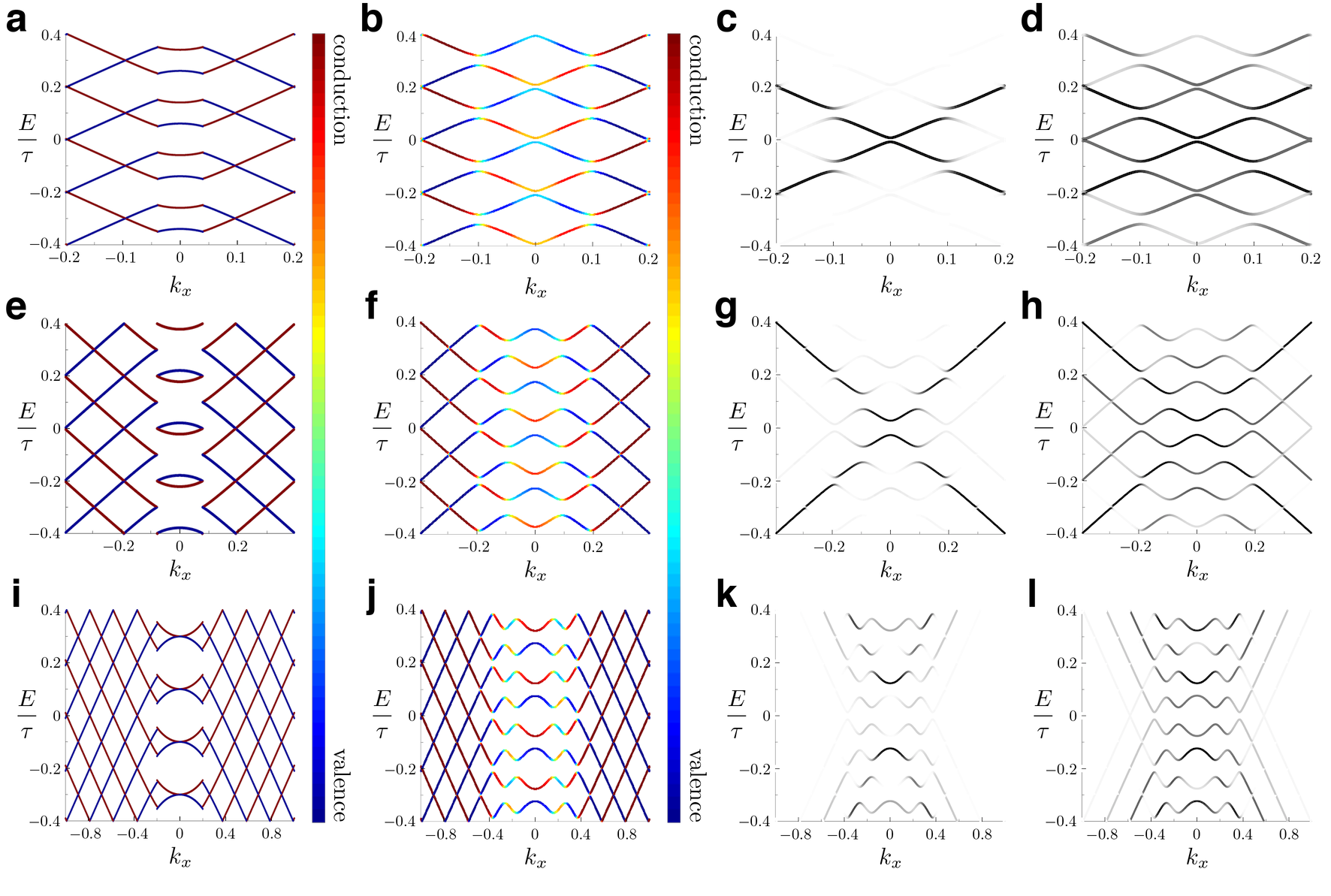}
\caption{
{\bf Quasienergy dispersion of the Floquet states in the continuum limit at an intermediate frequency with three representative ratios to the radius of the nodal helix.}
Specifically, $\Omega/\tau =16 A/\pi$ ({\bf a}\mbox{-}{\bf d}), $8 A/\pi$ ({\bf e}\mbox{-}{\bf h}), and $A$ ({\bf i}\mbox{-}{\bf l}) with $\Omega/\tau=0.2$.
({\bf a}, {\bf e}, {\bf i}) Quasienergy dispersion of the time Wannier-Stark ladder eigenstates.
({\bf b}, {\bf f}, {\bf j}) Quasienergy dispersion of the full Floquet states.
({\bf c}, {\bf g}, {\bf k}) Local density of Floquet states via the Gaussian probe weight with the standard deviation $\sigma$ being $1/2$.
({\bf d}, {\bf h}, {\bf l}) Local density of Floquet states via the Gaussian probe weight with the standard deviation $\sigma$ being $3/2$.
As one can see, at $\Omega/\tau =A$, the quasienergy dispersion of the Floquet states cannot be understood in terms of the simple overlapping Floquet copies of the Dirac dispersion.
\label{fig:Intermediate}
}
\end{center}
\end{figure*}

For simplicity, let us take the continuum limit of graphene or the surface of a topological insulator under radiation, which has the same Hamiltonian expressed in the form of Eq.~\eqref{eq:H_k} with $g_{\bf k}$ replaced by $g_{\bf q}=\tau [(q_x-A\cos{\kappa})+i(q_y-A\sin{\kappa})]$ with ${\bf q}=({\bf q}_\perp,\kappa)$ and ${\bf q}_\perp=(q_x,q_y)={\bf k}_\perp-{\bf K}_{\rm Dirac}$ being the momentum measured from the Dirac node.
The quasienergy of the time WSL eigenstates in this limit can be given analytically by Eq.~\eqref{eq:E_WSL2} with $\bar{\epsilon}_\pm({\bf q}_\perp)=\pm\frac{2\tau A}{\pi}(1+r)E{\left(\frac{2\sqrt{r}}{1+r}\right)}$, where $E(k)$ is the complete elliptic integral of the second kind and $r=\rho/A$ with $\rho=|{\bf q}_\perp|$.
Considering the overall shape of the elliptic integral, the quasienergy dispersion of the time WSL eigenstates can be understood as being roughly quadratic and linear inside and outside the projected nodal helix, respectively, with the relative topological shift by $\Omega/2$. 
Tuning the ratio between $A$ and $\Omega/\tau$ can create an interesting series of patterns in the quasienergy dispersion of the Floquet states.

Figure~\ref{fig:Intermediate} shows the quasienergy dispersion of the Floquet states at an intermediate frequency with three representative ratios to the radius of the nodal helix, say, $\Omega/\tau =16 A/\pi$, $8 A/\pi$, and $A$.
At $\Omega/\tau =16 A/\pi$, the quasienergy dispersion can be rather well understood in terms of the simple overlapping Floquet copies of the Dirac dispersion with slight gap opening whenever different bands cross each other.
Such a understanding is still possible at $\Omega/\tau =8 A/\pi$ despite some irregularities.  
At $\Omega/\tau =A$, however, it is no longer possible to do so. 
This change of patterns is due to the intricate interplay between the quasienergy dispersion inside and outside the projected nodal helix having the relative topological shift by $\Omega/2$.

To compare with the actual experimental data obtained via Tr-ARPES, it is convenient to compute the ``local density of Floquet states'' (LDOFS), which is analogous to the local density of states under a DC electric field~\cite{Lee2015, Kim2016}.  
 Mathematically, the LDOFS is defined as
\begin{align}
\rho_{\rm loc}(\omega,{\bf k}_\perp) = \sum_{\alpha,n} {\cal W}_{\rm probe}(n) \langle \tilde{\psi}^n_{\alpha,{\bf k}_\perp} | \tilde{\psi}^n_{\alpha,{\bf k}_\perp} \rangle \delta(\omega-\epsilon_{\alpha,{\bf k}_\perp}) ,
\end{align}
where the ``probe weight'' ${\cal W}_{\rm probe}(n)$ is the Gaussian-like localized weight indicating the probability for an electron to exist in the $n$-th time-conjugate Fourier component state after interacting with the probing light.
For example, ${\cal W}_{\rm probe}(n) \propto e^{-(n-\bar{n})^2/2\sigma^2}$ with $\bar{n}$ and $\sigma$ being the mean and the standard deviation, respectively.
Note that $|\tilde{\psi}^n_{\alpha,{\bf k}_\perp}\rangle$ is the $n$-th time-conjugate Fourier component of the $\alpha$-th Floquet state at ${\bf k}_\perp$ with the quasienergy eigenvalue $\epsilon_{\alpha,{\bf k}_\perp}$.
It is found that the quasienergy dispersion shown in the middle panels of Fig.~\ref{fig:Intermediate} [specifically, ({\bf h})] is quite consistent with the experimental data obtained by Wang et al. via Tr-ARPES~\cite{Wang2013}.

{\bf Discussion}

In this work, we have shown that a Floquet topological semimetal with nodal helix can be generated by irradiating graphene or the surface of a topological insulator with circularly polarized light.
The dynamics of such a Floquet topological semimetal is described by the TSH, where an effective electric field is applied along the axis of the nodal helix.
In the case of irradiated graphene, the TSH can host a Floquet topological insulator and a weakly driven Floquet topological semimetal with nodal helix in the high and low frequency limits, respectively. 
Importantly, in the low frequency limit, the $\pi$ shift of the Zak phase inside the projected nodal helix is predicted to generate the topological discontinuity along the projected nodal helix in the momentum spectrum of the Floquet states.
At intermediate frequency, this topological discontinuity can create an interesting change of patterns in the quasienergy dispersion of the Floquet states, which can be experimentally measured via Tr-ARPES.

We conclude this work by discussing the possibility of observing the topological surface flat band~\cite{Burkov2011}, also known as the drumhead surface state, at the boundary of the Floquet lattice.
Theoretically, the Floquet topological semimetal with nodal helix should have exactly the same topological surface flat band at the boundary as the usual topological semimetal with nodal line.
A question is what the boundary of the Floquet lattice means physically. 
The Floquet lattice site is labeled by the time-conjugate Fourier index denoting the $n$-th harmonics of the driving frequency. 
We believe that the boundary of the Floquet lattice can be interpreted as the maximum or minimum cutoff harmonics, up to which electrons can respond to the driving electric field.  
If so, the topological surface flat band can be observed near this cutoff frequency.

{\bf Methods}


{\bf High-frequency effective Hamiltonian of irradiated graphene.}
In the high frequency limit, the TSH can be expanded up to the order of $1/\Omega$ as follows:
\begin{align}
{\cal H}_{{\rm eff},\nu}= \nu\Omega \hat{I} +{\bf d}_{{\bf k}_\perp} \cdot \boldsymbol{\sigma} ,
\label{eq:H_eff2}
\end{align}
describing the effective dynamics of the $\nu$-th pair of the quasienergy band. 
For convenience, let us rewrite the results obtained in Eq.~\eqref{eq:H_eff}: 
\begin{align}
d_{{\bf k}_\perp,+}&=d_{{\bf k}_\perp,x} +i d_{{\bf k}_\perp,y} = \Gamma_0^*({\bf k}_\perp) ,
\label{eq:d_k+}
\\
d_{{\bf k}_\perp,-}&=d_{{\bf k}_\perp,x} -i d_{{\bf k}_\perp,y} = \Gamma_0({\bf k}_\perp) ,
\label{eq:d_k-}
\\
d_{{\bf k}_\perp,z}&=\sum_{n=1}^\infty \frac{|\Gamma_n({\bf k}_\perp)|^2-|\Gamma_{-n}({\bf k}_\perp)|^2}{n\Omega}, 
\label{eq:d_kz}
\end{align}
where
\begin{align}
\Gamma_n({\bf k}_\perp)=e^{i\phi_0({\bf k}_\perp)} \int_0^{2\pi} \frac{d\kappa}{2\pi} g_{\bf k} e^{i n\kappa} 
\label{eq:Gamma_n}
\end{align}
with $\phi_0({\bf k}_\perp)=\phi({\bf k}_\perp,\kappa=0)$.

In the nearest-neighbor tight-binding model of graphene,
\begin{align}
g_{\bf k}=-\tau \sum_{j=1}^3 e^{i{\bf k}^{\rm Pei\mbox{-}sub}_\perp \cdot {\bf c}_j} ,
\end{align}
where ${\bf k}_\perp^{\rm Pei\mbox{-}sub}=(k_x-A\cos{\kappa},k_y-A\sin{\kappa})$ with ${\bf c}_1=(-\sqrt{3}/2, -1/2)$, ${\bf c}_2=(\sqrt{3}/2, -1/2)$, and ${\bf c}_3=(0,1)$ in units of lattice spacing, which is set to be unity.
After some algebra, one can show that
\begin{align}
g_{\bf k} 
= -\tau \sum_{j=1}^3 e^{i{\bf k}_\perp \cdot {\bf c}_j} e^{-i A \cos{(\kappa-\theta_j)} } ,
\end{align}
where $\cos{\theta_j}={\bf c}_{j,x}$ and $\sin{\theta_j}={\bf c}_{j,y}$.
Then, Eq.~\eqref{eq:Gamma_n} can be rewritten as follows:
\begin{align}
\frac{\Gamma_n({\bf k}_\perp)}{e^{i\phi_0({\bf k}_\perp)}} 
&= -\tau \sum_{j=1}^3 e^{i{\bf k}_\perp \cdot {\bf c}_j} 
\int_0^{2\pi} \frac{d\kappa}{2\pi} e^{i n\kappa} e^{-i A \cos{(\kappa-\theta_j)} } 
\nonumber \\
&= -\tau i^n J_n(-A) \sum_j e^{i{\bf k}_\perp \cdot {\bf c}_j} e^{i n\theta_j} ,
\label{eq:Gamma_n2}
\end{align}
where the definition of the Bessel function has been used: $J_n(z)=\int \frac{d \phi}{2\pi i^n} e^{in\phi}e^{iz\cos{\phi}}$.

We are interested in the modulus square of $\Gamma_n({\bf k}_\perp)$ for $n \neq 0$:
\begin{align}
|\Gamma_n({\bf k}_\perp)|^2
= \tau^2 J^2_n(A)  \Big| \sum_{j=1}^3 e^{i{\bf k}_\perp \cdot {\bf c}_j} e^{i n\theta_j} \Big|^2 .
\label{eq:Gamma_n_mod_square}
\end{align}
Meanwhile, for $n=0$, $\Gamma_0({\bf k}_\perp)$ itself is important:
\begin{align}
\Gamma_0({\bf k}_\perp)
= -\tau J_0(A) e^{i\phi_0({\bf k}_\perp)} \sum_{j=1}^3 e^{i{\bf k}_\perp \cdot {\bf c}_j} ,
\label{eq:Gamma_0}
\end{align}
where we have used $J_n(-z)=(-1)^n J_n(z)$.
By using these expressions, one can rewrite Eq.~\eqref{eq:d_kz} as follows:
\begin{align}
d_{{\bf k}_\perp,z} = 
\frac{\tau^2}{\Omega}\sum_{n=1}^\infty \frac{J_n^2(A)}{n} {\cal F}_n({\bf k}_\perp) ,
\label{eq:d_kz2}
\end{align}
where we have used $J_{-n}(z)=(-1)^n J_n(z)$, and
\begin{align}
{\cal F}_n({\bf k}_\perp)
&=
\Big| \sum_{j=1}^3 e^{i{\bf k}_\perp \cdot {\bf c}_j} e^{i n\theta_j} \Big|^2
-\Big| \sum_{j=1}^3 e^{i{\bf k}_\perp \cdot {\bf c}_j} e^{-i n\theta_j} \Big|^2 
\nonumber \\
&=
2i \sum_{j \neq k} \sin{(n \theta_{jk})} e^{i {\bf k}_\perp \cdot {\bf c}_{jk}} ,
\label{eq:cal_F}
\end{align}
where $\theta_{jk}=\theta_j-\theta_k$ and ${\bf c}_{jk}={\bf c}_j-{\bf c}_k$.
Now, it is important to note that ${\bf c}_{jk}$ are actually the displacement vectors connecting between next-nearest-neighboring sites in graphene.
That is, there are six displacement vectors: $\boldsymbol{\eta}_1={\bf c}_{13}$, $\boldsymbol{\eta}_2={\bf c}_{23}$, $\boldsymbol{\eta}_3 = {\bf c}_{21}$,  $\boldsymbol{\eta}_4={\bf c}_{31}$, $\boldsymbol{\eta}_5={\bf c}_{32}$, and $\boldsymbol{\eta}_6={\bf c}_{12}$.
Also noting that $\theta_{12}=\theta_{23}=\theta_{31}=-\theta_{21}=-\theta_{32}=-\theta_{13}=-2\pi/3$, one can rewrite Eq.~\eqref{eq:cal_F} as follows:
\begin{align}
{\cal F}_n({\bf k}_\perp)=
2i \sin{\left(\frac{2n\pi}{3}\right)}\sum_{j=1}^6 (-1)^j e^{i {\bf k}_\perp \cdot \boldsymbol{\eta}_j} .
\label{eq:cal_F2}
\end{align}
Finally, after collecting all the factors, $d_{{\bf k}_\perp,z}$ can be written as
\begin{align}
d_{{\bf k}_\perp,z} = 
i \lambda_{\rm eff}(A) \sum_{j=1}^6 (-1)^j e^{i {\bf k}_\perp \cdot \boldsymbol{\eta}_j} ,
\label{eq:d_kz_final}
\end{align}
where
\begin{align}
\frac{\lambda_{\rm eff}(A)}{\tau^2/\Omega} = 2 \sum_{n=1}^\infty \frac{J_n^2(A)}{n} \sin{\left( \frac{2n\pi}{3} \right)} .
\label{eq:lambda_eff}
\end{align}

Also, by using Eq.~\eqref{eq:Gamma_0}, $d_{{\bf k}_\perp,\pm}$ can be written as
\begin{align}
d^*_{{\bf k}_\perp,+}=d_{{\bf k}_\perp,-}
= - \tau_{\rm eff}(A) e^{i\phi_0({\bf k}_\perp)} \sum_{j=1}^3 e^{i{\bf k}_\perp \cdot {\bf c}_j} ,
\label{eq:d_k_pm_final}
\end{align}
where $\tau_{\rm eff}(A)=\tau J_0(A)$.
Note that the phase factor $e^{i\phi_0({\bf k}_\perp)}$ is not important since it can be removed via an appropriate gauge transformation.

{\bf Quasienergy gap near the projected nodal helix at low frequency.}
The quasienergy gap opens up when two different time WSL eigenstates with one from the conduction band and the other from the valence band are mixed together.
The mixing matrix element between these two time WSL eigenstates is given by
\begin{align}
{\cal M}_{nn^\prime}({\bf k}_\perp) = \frac{\Omega}{2} \int^{2\pi}_0 \frac{d\kappa}{2\pi} [\psi^{\rm WSL}_{+,n}({\bf k})]^*  \frac{\partial \phi({\bf k})}{\partial \kappa} \psi^{\rm WSL}_{-,n^\prime}({\bf k}) ,
\label{eq:mix_mat_elem}
\end{align}
since ${\cal A}^{\pm \mp}_{\kappa}({\bf k})=\frac{1}{2}\frac{\partial}{\partial \kappa} \phi({\bf k})$.
Also, 
\begin{align}
\psi^{\rm WSL}_{\pm,n} ({\bf k}) = e^{ 
-\frac{i}{\Omega} \int^\kappa_0 d\kappa^\prime 
\left[ {\cal E}^{\rm WSL}_{\pm,n} ({\bf k}_\perp) -\epsilon_\pm({\bf k}^\prime) -\frac{\Omega}{2}\frac{\partial}{\partial \kappa}\phi({\bf k}^\prime) \right]  
} ,
\label{eq:psi_WSL_graphene}
\end{align}
since ${\cal A}^{\pm \pm}_{\kappa}({\bf k})=\frac{1}{2}\frac{\partial}{\partial \kappa} \phi({\bf k})$.

Interestingly, all the terms dependent on the Zak phase completely cancel in the mixing matrix element, which can be thus written in a single formula regardless of whether ${\bf k}_\perp$ is located inside or outside the projected nodal helix:
\begin{align}
{\cal M}_{nn^\prime}({\bf k}_\perp) = \frac{\Omega}{2} \int^{2\pi}_0 \frac{d\kappa}{2\pi}  \frac{\partial \phi({\bf k})}{\partial \kappa} 
e^{-i \Delta n \kappa} e^{\frac{i}{\Omega} \int^\kappa_0 d\kappa^\prime \left[ \Delta \epsilon({\bf k}^\prime) -\Delta \bar{\epsilon} \right] } ,
\label{eq:mix_mat_elem2}
\end{align}
where $\Delta n=n-n^\prime$, $\Delta \epsilon({\bf k})= \epsilon_+({\bf k})-\epsilon_-({\bf k})$, and $\Delta \bar{\epsilon}= \bar{\epsilon}_+({\bf k}_\perp)-\bar{\epsilon}_-({\bf k}_\perp)$.
By inspecting the form of the above integral, one can find that the most important contribution comes from the region, where $\phi({\bf k})$ changes significantly as a function of $\kappa$.

It can be shown from Eq.~\eqref{eq:insta_eigenstates} that $\phi({\bf k})$ changes significantly around the critical value of $\kappa$, $\kappa_{\rm c}({\bf k}_\perp)$, where the nodal helix is the closest to the constant ${\bf k}_\perp$ line.
For convenience, let us define $f(\kappa)=\frac{\partial}{\partial \kappa}\phi({\bf k})$, which is written as a function of $\kappa$ with the ${\bf k}_\perp$ dependence being implicit. 
Now, one can expand $f(\kappa)$ around $\kappa_{\rm c}$; $f(\kappa) \simeq f_{\rm c} +f^\prime_{\rm c} \Delta\kappa +\frac{f^{\prime\prime}_{\rm c}}{2} (\Delta\kappa)^2$, where $f_{\rm c}=f(\kappa=\kappa_{\rm c})$, $f^\prime_{\rm c}=f^\prime(\kappa=\kappa_{\rm c})$, $f^{\prime\prime}_{\rm c}=f^{\prime\prime}(\kappa=\kappa_{\rm c})$, and $\Delta\kappa=\kappa-\kappa_{\rm c}$.
Since $f^\prime_{\rm c}=0$ due to the very definition of $\kappa_{\rm c}$, one can approximate $f(\kappa)$ as follows:
\begin{align}
f(\kappa) \simeq f_{\rm c} e^{\frac{f^{\prime\prime}_{\rm c}}{2 f_{\rm c}} (\Delta\kappa)^2} .
\label{eq:f_kappa_expand}
\end{align}
Under this expansion scheme, one can then rewrite Eq.~\eqref{eq:mix_mat_elem2} as follows:
\begin{align}
\frac{{\cal M}_{nn^\prime}}{\Omega f_{\rm c}/2} 
\simeq
\int^{2\pi}_0 \frac{d\kappa}{2\pi} 
e^{\frac{f^{\prime\prime}_{\rm c}}{2 f_{\rm c}} (\Delta\kappa)^2} 
e^{-i \Delta n \kappa} e^{\frac{i}{\Omega} \int^\kappa_0 d\kappa^\prime \left[ \Delta \epsilon({\bf k}^\prime) -\Delta \bar{\epsilon} \right] } .
\label{eq:mix_mat_elem3}
\end{align}
Note that the ${\bf k}_\perp$ dependence is not explicitly written from this forward unless it is necessary.

In the limit of low frequency, the last exponential term in Eq.~\eqref{eq:mix_mat_elem3} fluctuates wildly unless its argument is nearly zero. 
To take care of this wildly fluctuating term, let us first rewrite Eq.~\eqref{eq:mix_mat_elem3} as follows:
\begin{align}
\frac{{\cal M}_{nn^\prime}} {\Omega f_{\rm c}/2} 
\simeq 
{\cal C}
\int^{2\pi}_0 \frac{d\kappa}{2\pi} 
e^{\frac{f^{\prime\prime}_{\rm c}}{2 f_{\rm c}} (\Delta\kappa)^2} 
e^{-i \Delta n \Delta\kappa} e^{\frac{i}{\Omega} \int^\kappa_{\kappa_{\rm c}} d\kappa^\prime \left[ \Delta \epsilon({\bf k}^\prime) -\Delta \bar{\epsilon} \right] } ,
\label{eq:mix_mat_elem4}
\end{align}
where ${\cal C}=e^{-i \Delta n \kappa_{\rm c}} e^{\frac{i}{\Omega} \int^{\kappa_{\rm c}}_0 d\kappa^\prime \left[ \Delta \epsilon({\bf k}^\prime) -\Delta \bar{\epsilon} \right]}$ is just a constant phase factor.
Then, we expand the argument of the last exponential term up to the linear of order of $\Delta\kappa$ as follows:
\begin{align}
\frac{{\cal M}_{nn^\prime}} {\Omega f_{\rm c}/2} 
\simeq 
{\cal C}
\int^{2\pi}_0 \frac{d\kappa}{2\pi} 
e^{\frac{f^{\prime\prime}_{\rm c}}{2 f_{\rm c}} (\Delta\kappa)^2} 
e^{-i \left[ \Delta n -\frac{1}{\Omega}  (\Delta \epsilon_{\rm c} -\Delta \bar{\epsilon}) \right] \Delta\kappa} ,
\label{eq:mix_mat_elem5}
\end{align}
where $\Delta \epsilon_{\rm c}=\Delta \epsilon({\bf k}_\perp, \kappa=\kappa_{\rm c})$.

Then, as a final approximation, we take the upper and lower limits of the integral to be plus and minus infinity and perform the Gaussian integration with respect to $\Delta\kappa$:
\begin{align}
\frac{{\cal M}_{nn^\prime}} {\Omega f_{\rm c}/2} 
&\simeq 
{\cal C}
\int^{\infty}_{-\infty} \frac{d\kappa}{2\pi} 
e^{-\alpha (\Delta\kappa)^2} 
e^{i \beta\Delta\kappa} 
\nonumber \\
&=
\frac{{\cal C}}{2\pi} \sqrt{\frac{\pi}{\alpha}} e^{-\frac{\beta^2}{4\alpha}} ,
\label{eq:mix_mat_elem6}
\end{align}
where $\alpha=-f^{\prime\prime}_{\rm c}/2f_{\rm c}$ and $\beta=\frac{1}{\Omega}  (\Delta \epsilon_{\rm c} -\Delta \bar{\epsilon}) -\Delta n$.
Note that the above Gaussian integration is valid since $\alpha>0$ as shown by an explicit calculation.

We are interested in the quasienergy gap at zero quasienergy, or any other quasienergy value where the quasienergy of the time WSL eigenstate from the conduction band matches exactly that from the valence band.
Mathematically, this means that $\Delta \bar{\epsilon}+\Omega \Delta n=0$.
In other words, the diagonal matrix element is exactly the same for both of the time WSL eigenstates from the conduction and valence bands.
In this situation, the quasienergy gap is given simply by twice the absolute value of the mixing matrix element:
\begin{align}
{\cal E}_{\rm gap} & = 2 |{\cal M}_{\Delta n = -\Delta \bar{\epsilon}/\Omega}|
\nonumber \\
&\simeq
\frac{\Omega f_{\rm c}}{2\pi}
\sqrt{\frac{\pi}{\alpha}}
e^{-\frac{(\Delta \epsilon_{\rm c})^2}{4\alpha\Omega^2}  }.
\label{eq:E_gap}
\end{align}

Finally, in the continuum limit, the graphene Hamiltonian is taken to be linear in the vicinity of a given Dirac node, i.e., $g_{\bf q}=\tau[(q_x-A\cos{\kappa})+i(q_y-A\sin{\kappa})]$, where ${\bf q}_\perp=(q_x,q_y)={\bf k}_\perp-{\bf K}_{\rm Dirac}$ is the momentum measured from the Dirac node.
Here, $f(\kappa)=\frac{\partial}{\partial \kappa} \phi(q_x,q_y,\kappa)$ can be obtained analytically:
\begin{align}
f({\kappa}) =
\frac{1-\frac{\rho}{A}\cos{(\kappa-\theta)}}{1-(\frac{\rho}{A})^2-2\frac{\rho}{A}\cos{(\kappa-\theta)}} ,
\label{eq:f_kappa_continuum}
\end{align}
where $\rho=\sqrt{q_x^2+q_y^2}$ and $\tan{\theta}=q_y/q_x$.
The above formula can be derived by noting that $g_{\bf q}=\rho e^{i\theta} -A e^{i\kappa}$.
By using this formula, one can determine $f_{\rm c}=\frac{A}{A-\rho}$ and $\alpha=\frac{\rho(A+\rho)}{2(A-\rho)^2}$ with $\kappa_{\rm c}=\theta$. 
Also, one can show $\Delta \epsilon_{\rm c}=2|g_{({\bf q}_\perp,\kappa_{\rm c})}|=2\tau| A-\rho |$ in the continuum limit.
Inserting $f_{\rm c}$, $\alpha$, and $\Delta \epsilon_{\rm c}$ into Eq.~\eqref{eq:E_gap}, we arrive at the following formula for the quasienergy gap in the continuum limit:
\begin{align}
{\cal E}_{\rm gap} 
\simeq
\frac{\Omega A}{\sqrt{2\pi\rho(A+\rho)}}
e^{-\frac{2(A-\rho)^4}{\rho(A+\rho)(\Omega/\tau)^2}  } .
\end{align}

It is important to note that, due to the exponential suppression term, the quasienergy gap near the projected nodal helix vanishes much faster than $\Omega$ as soon as $\rho$ becomes separated from $A$ by roughly more than $\sqrt{A\Omega/\tau}$.
Consequently, in the low frequency limit, the time WSL eigenstates become more and more sharply defined with the correspondingly pronounced topological discontinuity along the projected nodal helix.

{\bf Abelian weight of the full Floquet states.}
Here, we provide some details on how to compute the Abelian weight of the full Floquet states obtained by diagonalizing the Floquet Hamiltonian. 
In other words, we would like to decompose the full Floquet states, $|\psi_\alpha({\bf k})\rangle$, in terms of the conduction/valence time WSL eigenstates, $\psi^{\rm WSL}_{\pm,n} ({\bf k})$, obtained by ignoring the off-diagonal elements of the Berry connection. 
Mathematically, the Abelian weight of the $\alpha$-th full Floquet states is defined as follows:
\begin{align}
{\cal W}_{\pm,\alpha}({\bf k}_\perp) &= \sum_n \left|   \int^{2\pi}_0 \frac{d\kappa}{2\pi} \psi^*_{\pm,\alpha}({\bf k}) \psi^{\rm WSL}_{\pm,n}({\bf k}) \right|^2 
\nonumber \\
&= \int^{2\pi}_0 \frac{d\kappa}{2\pi} \left|\psi_{\pm,\alpha}({\bf k})\right|^2 ,
\label{eq:Abelian_weight}
\end{align}
where $\psi_{\pm,\alpha}({\bf k})$ is the $\pm$, or conduction/valence component of $|\psi_\alpha({\bf k})\rangle$.
The second line is obtained due to the completeness of the time WSL eigenstates. 
Also, ${\cal W}_{+,\alpha}+{\cal W}_{-,\alpha}=1$ due to the normalization of the Floquet states.
Therefore, the Abelian weight is nothing but the projection weight of the full Floquet states onto the instantaneous energy eigenstate basis.

Now, suppose that one obtains the full Floquet states by diagonalizing the Floquet Hamiltonian $\hat{H}_{\rm F}$ expressed in terms of the Floquet and sublattice index, not the TSH expressed in terms of time and the instantaneous energy eigenvalue index.
That is, 
\begin{align}
|\psi_\alpha({\bf k})\rangle=\sum_{m} e^{-im\kappa} 
\left(
\begin{array}{c}
a_{\alpha,m}({\bf k}_\perp) \\
b_{\alpha,m}({\bf k}_\perp) 
\end{array} 
\right) ,
\end{align}
where $a_{\alpha,m}$ and $b_{\alpha,m}$ denote the amplitude of the Floquet state in each sublattice basis.
To compute the Abelian weight in terms of this representation, one needs to perform a unitary transformation of this Floquet state to the instantaneous energy eigenstate basis. 
Mathematically,
\begin{align}
\psi_{\pm,\alpha}({\bf k})= \langle \pm | U^\dagger({\bf k}) |\psi_\alpha({\bf k})\rangle ,
\end{align}
where $\langle \pm |=(1,0)$ and $(0,1)$, respectively.
The unitary transformation matrix is given by
\begin{align}
U({\bf k}) 
&=\frac{1}{\sqrt{2}}
\left(
\begin{array}{cc}
e^{-i\phi({\bf k})} & e^{-i\phi({\bf k})} \\
1 & -1
\end{array} 
\right) 
\nonumber \\
&=\frac{1}{\sqrt{2}}
\sum_n e^{-in\kappa}
\left(
\begin{array}{cc}
\varphi_n({\bf k}_\perp) & \varphi_n({\bf k}_\perp) \\
\delta_{n,0} & -\delta_{n,0}
\end{array} 
\right) ,
\end{align}
where $\varphi_n({\bf k}_\perp)$ is the $n$-th time-conjugate Fourier component of $e^{-i\phi({\bf k})}$.
Then, after plugging all the above expressions into Eq.~\eqref{eq:Abelian_weight} and performing the time integration, one can obtain the following result:
\begin{align}
{\cal W}_{\pm,\alpha} = \frac{1}{2} \pm \sum_{n,m} {\rm Re}(\varphi_{n} a^*_{\alpha,m} b_{\alpha,m-n}) ,
\end{align}
where the ${\bf k}_\perp$ argument is not explicitly written for simplicity.

{\bf Acknowledgements}

The authors are grateful to Sutirtha Mukherjee and Changsuk Noh for their insightful comments.
The authors thank the KIAS Center for Advanced Computation (CAC) for providing computing resources.

\end{document}